# Service-Fingerprinting mittels Fuzzing

Michael Hanspach[1] · Ralf Schumann[1]

Stefan Schemmer[2] · Sebastian Vandersee[2]

[1]Fachhochschule der Wirtschaft FHDW, Bergisch Gladbach

[2]rt-solutions.de GmbH, Köln

## Zusammenfassung

Für die Durchführung effektiver Penetrationstests ist die Identifizierung von Diensten (Services) und Applikationen auf den Zielsystemen, das sogenannte Service-Fingerprinting, von zentraler Bedeutung. Ziel dieses Beitrags ist es, mögliche Verbesserungspotentiale bestehender Fingerprinting-Tools zu beleuchten. Dies soll durch Einsatz von Mutation-Based Fuzzing zwecks einfacher und automatischer Erzeugung von Fingerprints erreicht werden. Durch Versenden der so präparierten Anfragen an bestimmte Dienste und Dienstversionen sollen subtile Unterschiede in den Antwortnachrichten erfasst werden, die Identifizierungs- und Unterscheidungsmerkmale bilden.

Für den Nachweis der Praxistauglichkeit des Fuzzing-basierten Ansatzes und zum Vergleich mit bereits bestehenden Methoden und Tools wird eine Implementierung zum Fingerprinting von FTP-Servern vorgestellt.

Im Ergebnis dieses Beitrags wird aufgezeigt, dass Fuzzing als Grundlage für das Service-Fingerprinting geeignet und sogar in Teilen den bestehenden Ansätzen, insbesondere bei der Unterscheidung von Versionen einzelner Dienste, überlegen ist.

## 1 Einleitung

Im Hinblick auf die IT-Compliance von Unternehmen kommt der Informationssicherheit eine herausragende Bedeutung zu. In diesem Kontext stellen Penetrationstests eine wichtige Maßnahme zur Informationssicherheit dar. Ziel solcher Tests ist es, unter Verwendung verschiedener Angriffsvektoren die Sicherheit von Netzwerken und Systemen sowohl automatisiert als auch manuell auf potentielle Schwachstellen hin zu untersuchen und diese zu bewerten. Je nach Bedrohungsszenario werden die Analysen aus unterschiedlichen Sichtweisen durchgeführt, namentlich Blackbox- oder Whitebox-Tests. Penetrationstests bilden folglich eine wichtige Entscheidungsgrundlage zur Ergreifung geeigneter Maßnahmen, um identifizierte Schwachstellen zu beheben oder zumindest zu reduzieren.

Die initiale Phase eines Penetrationstests stellt üblicherweise die grundlegende Informationssammlung über Basiseigenschaften der Zielsysteme dar. Dies umfasst insbesondere das sogenannte Fingerprinting, d.h. die Identifizierung von Betriebssystemen („OS-Fingerprinting") und Anwendungen bzw. Diensten („Service-Fingerprinting") einschließlich genauer Versionsstände. Je genauer sich in dieser Phase Aussagen über die Zielsysteme ermitteln lassen, desto effizienter lässt sich das Vorgehen für weitere Tests und Analysen festlegen. Ein zuverlässiges Fingerprinting hat demnach hohe Praxisrelevanz für Penetrationstests. Ziel des vorliegenden Papers ist es, eine neue Fingerprinting-Methode zu entwickeln, die im Vergleich zu bestehenden Fingerprinting-Verfahren eine genauere Erkennung von Diensten ermöglicht und gleichzeitig flexibel auf neue Dienste und Versionsstände reagiert.

## 2 Bestehende Verfahren

Verwandte Arbeiten im Bereich OS- und Service-Fingerprinting sollen im Folgenden kurz dargestellt und vom Thema dieses Papers abgegrenzt werden.

[Lippmann et al.] präsentieren eine Studie über passives OS-Fingerprinting, bei der systematisch verschiedene Klassen von Betriebssystemen gebildet werden. Im Gegensatz hierzu beschäftigt sich das vorliegende Paper mit aktivem Fingerprinting, d.h. es werden initiativ Anfrage-Nachrichten erzeugt und versendet, um die zugehörigen Antwort-Nachrichten systematisch zu analysieren.

[Sarraute und Burroni] berichten über ein fortgeschrittenes OS-Fingerprinting-Verfahren, dass auf neuronalen Netzwerken basiert. Im Gegensatz zum vorliegenden Paper verwenden Sarraute und Burroni jedoch keine systematisch erzeugte Anfragemenge. Durch die systematische Erzeugung von Anfrage-Nachrichten mittels Fuzzing können prinzipiell beliebige Unterschiede im Antwort-Verhalten von Betriebssystemen und Diensten erkannt werden.

[Gagnon et al.] schlagen eine hybride Vorgehensweise bzgl. des OS-Fingerprintings vor, die passives und aktives Fingerprinting miteinander verbindet. Dabei verwenden Gagnon et al. wissensbasierte Systeme, die nur dann aktive Fingerprinting-Verfahren einsetzen, wenn dieses notwendig ist, um auf der Basis bereits vorhandenen Informationen eine weitere Eingrenzung vorzunehmen. Im Unterschied dazu werden im vorliegenden Paper nur aktive Fingerprinting-Mechanismen verwendet und es wird bewusst keine Einschränkung der Anfrage-Menge vorgenommen, um möglichst viele verschiedene Betriebssysteme und Dienste voneinander abgrenzen zu können.

[Caballero et al.] beschreiben eine erste Studie über Fingerprinting auf Basis automatisierter Anfrage-Nachrichten. Anders als im vorliegenden Paper werden dabei Methoden des Machine Learning eingesetzt. Der Einsatz von Fuzzing hat gegenüber Machine Learning den Vorteil, das Anfrage-Antwort-Verhalten verschiedener Systeme (auch zukünftiger) möglichst detailliert abzubilden. Der Nachteil ist jedoch, dass erheblich mehr Anfrage-Nachrichten versendet werden müssen.

[Greenwald und Thomas] beschreiben, wie man den Netzwerkverkehr eines aktiven Fingerprinting-Verfahren verbergen kann. Im Unterschied zum vorliegenden Paper werden nur 1 bis 3 Anfrage-Nachrichten zum Fingerprinting verwendet. Während diese Maßnahmen die Detektion des Fingerprintings erschweren können, sind diese jedoch nicht mit einer systematischen Analyse auf Basis von großen Mengen automatisiert erzeugter Anfrage-Nachrichten vereinbar. Allerdings wäre es durchaus denkbar, auf Basis der hier durchgeführten Analyse eine kleine Menge spezifischer Anfrage-Nachrichten auszuwählen, die für das OS- und Service-Fingerprinting von besonderer Relevanz ist.

Eine aktive Fingerprinting-Technik besteht darin, Statuscodes von Anwendungsprotokollen zu nutzen. Insbesondere Statuscodes als Reaktion auf Fehlersituationen unterscheiden sich oft geringfügig zwischen Dienstimplementierungen und teilweise auch zwischen verschiedenen Versionen einzelner Dienste. Auch die Anwesenheit oder Abwesenheit von Statuscodes kann Hinweise auf die Identität eines Dienstes liefern. Beispiele für Fingerprinting-Software, die sich dieser Techniken bedient, sind hmap [hmap] (zur Identifizierung von Webservern) und FTPMap [FTPmap] (zur Identifizierung von FTP-Servern). Die Genauigkeit dieser Form des aktiven Fingerprintings hängt entscheidend von der Menge der verwendeten Anfragenachrichten und der Unterscheidbarkeit der Antworten ab. Mit Hilfe von Mutation-Based Fuzzing soll die Menge der Anfragenachrichten erheblich vergrößert werden, wie im Folgenden beschrieben.

## 3 Service-Fingerprinting mittels Fuzzing

Um die Suche nach Unterscheidungsmerkmalen zu automatisieren, wird das sogenannte Mutation-Based Fuzzing eingesetzt, ein Verfahren, das normalerweise in der Softwareentwicklung verwendet wird, um systematische Fehlersuche und –analyse zu betreiben. Im vorliegenden Kontext wird Mutation-Based Fuzzing jedoch dazu genutzt, systematisch Anfragenachrichten zu erzeugen. Die Antworten auf diese Anfragenachrichten werden für das Fingerprinting von Diensten herangezogen.

Somit entfällt die manuelle Suche nach Unterscheidungsmerkmalen. Zudem können durch die Automatisierung der Anfragenachrichtenerzeugung auch bisher unbekannte Unterscheidungsmerkmale identifiziert werden, die nicht oder nur sehr aufwendig durch eine manuelle Suche – etwa durch Quellcodeanalyse – entdeckt werden können. Ziel des Fuzzing-Ansatzes ist es, eine größere Fingerprinting-Genauigkeit zu erzielen. Dabei erlaubt es die automatische Erzeugung der Anfragenachrichten, flexibel Fingerprints für verschiedenste Protokolle zu erzeugen.

### 3.1 Vorgehensmodell

Das Vorgehensmodell des hier beschriebenen Ansatzes gliedert sich in zwei wesentliche Phasen.

Die *Laborphase* dient zur Erzeugung von Fingerprints zwecks Unterscheidung von Diensten. Die einzelnen Schritte der Laborphase sind in Abbildung 1 schematisch dargestellt und werden im Folgenden beschrieben.

Für das betreffende Anwendungsprotokoll wird zunächst eine Protokollanalyse durchgeführt, mit dem Ziel, die relevanten Protokollfelder zu identifizieren, um diese später mittels Fuzzing füllen zu können. Der Eingaberaum für das Fuzzing definiert sich damit durch jegliche Nachrichten, die über das Anwendungsprotokoll ausgetauscht werden können.

Im nächsten Schritt werden mittels Fuzzing Anfragenachrichten generiert. Diese werden im Anschluss an die jeweiligen Zieldienste versandt und die entsprechenden Antwortnachrichten, die die Fingerprints darstellen, aufgezeichnet und analysiert. Idealerweise werden nachfolgend im Rahmen einer Optimierung genau jene Anfragenachrichten ausgewählt, die bei den getesteten Diensten zu unterschiedlichen Fingerprints führen. So kann die Menge der Anfragenachrichten geeignet reduziert werden, ohne bei den getesteten Diensten an Genauigkeit zu verlieren. Da sich jeder Fingerprint auf eine bestimmte Menge von Anfragenachrichten bezieht und ein Vergleich zweier Fingerprints nur bei der gleichen Menge von Anfragenachrichten sinnvoll ist, müssen erneut Fingerprints der getesteten Dienste mit der reduzierten Menge von Anfragenachrichten erzeugt werden. Letztlich führt diese Optimierung dazu, dass das Fingerprinting während eines Penetrationstests schneller durchgeführt werden kann.

In der *Produktivphase* kommen die im Vorfeld generierten Anfragenachrichten und Fingerprints im Rahmen eines Penetrationstests zum Einsatz. So kann ein schneller Vergleich des Fingerprints eines Dienstes mit den Fingerprints bereits bekannter Dienste erfolgen. Der Vergleich zweier Fingerprints ergibt eine prozentuale Übereinstimmung, die bei hoher Übereinstimmung auf einen bestimmten Dienst oder eine bestimmte Dienstversion schließen lässt.

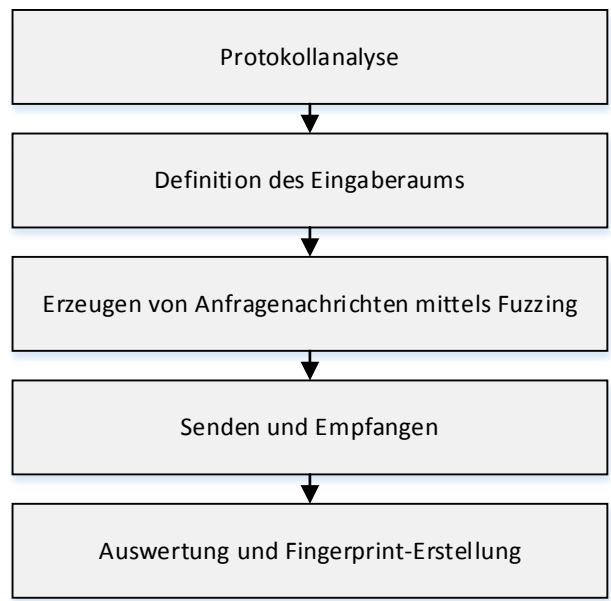

**Abb. 1:** Schritte der Laborphase

Das beschriebene 2-phasige Vorgehen ist dabei den Bedürfnissen der Praxis angepasst. Im Rahmen eines Penetrationstests, der typischerweise zeitlich sehr eingeschränkt realisiert wird, kann keine Erzeugung von Fingerprints erfolgen, zumal es sich häufig um Tests ohne Vorwissen über die Dienste handelt (sogenannte Blackbox-Tests). Zudem werden oft Produktivsysteme untersucht, deren Verfügbarkeit so gering wie möglich zu beeinträchtigen ist. Häufig sind Produktivsysteme auch zusätzlich mit vorgelagerten IPS (Intrusion Prevention Systems) geschützt, die ungewöhnlichen Netzwerkverkehr unterbinden, was das Fingerprinting und damit Penetrationstests u.U. erheblich erschwert oder verhindert. Daher ist es erforderlich, die Fingerprints im Vorfeld zu erzeugen, um danach Penetrationstests innerhalb eines kurzen Zeitraums durchführen zu können.

## 3.2 Auswahl von Fuzzing-Verfahren

Für die automatische Erzeugung von Anfragenachrichten existieren verschiedene Fuzzing-Methoden. Die einfachste Variante ist es, Bestandteile einer Anfragenachricht mittels eines Pseudozufallszahlengenerators zu füllen. Diese Methode bezeichnet man auch als Random Fuzzing. Problematisch ist jedoch, eine sinnvolle Größe für die Menge der zu erzeugenden Anfragenachrichten zu finden. Die Suche nach Unterscheidungsmerkmalen mittels Random Fuzzing kann u.U. lange Zeit in Anspruch nehmen, da sie nicht protokoll-spezifisch und damit nicht hinreichend zielführend ist. Zudem kann eine vollständige Erfassung von Unterscheidungsmerkmalen selbst nach langer Fuzzing-Laufzeit nicht garantiert werden.

Eine fortgeschrittenere Fuzzing-Methode ist das Mutation-Based Fuzzing, das die Grundlage für den hier beschriebenen Ansatz bildet. Beim Mutation-Based Fuzzing wird zunächst von gültigen Anfragenachrichten ausgegangen. Die gültigen Anfragenachrichten mutieren dann wiederholt, um ein möglicherweise unerwartetes Verhalten zu produzieren, das sich als Identifizierungsmerkmal eines Dienstes eignet. Auf diese Weise wird der Eingaberaum zielgerichtet eingeschränkt, was zu einer schnelleren Identifizierung von Unterscheidungsmerkmalen führt.

Bei der Erzeugung von Anfragenachrichten mittels Fuzzing entsteht schnell eine große Zahl von Anfragenachrichten, die in einem Penetrationstest nicht sinnvoll genutzt werden kann. Daher wird möglichst im Anschluss eine Auswahl jener Anfragenachrichten vorgenommen, die für das Fingerprinting die entscheidenden Anhaltspunkte liefern.

## 4 Implementierung

Der erläuterte Ansatz wurde im Rahmen einer Fingerprinting-Software für FTP-Server implementiert, ist für andere Protokolle jedoch grundsätzlich ähnlich realisierbar.

Entsprechend dem in Kapitel 3 erläuterten Vorgehensmodell wurde zunächst eine Protokollanalyse durchgeführt. Eine FTP-Sitzung besteht aus zwei getrennten TCP-Verbindungen, einer Kontrollverbindung für die Steuerung der Anwendung sowie einer Datenverbindung, über die die angeforderten Dateien transferiert werden. Ziel der Betrachtung ist ausschließlich die Kontrollverbindung des FTP-Servers, nicht jedoch die Datenverbindung, da die vom FTP-Server bereitgestellten Dateien für das Fingerprinting irrelevant sind. Als Eingaberaum dienen somit jegliche Nachrichten, die über die FTP-Kontrollverbindung ausgetauscht werden.

Anschließend wurden gültige Befehle typischer FTP-Server identifiziert. Dies umfasst insbesondere auch FTP-Befehle, die nicht standardisiert sind und daher ein gutes Unterscheidungsmerkmal bilden. Dabei wurden nur solche Befehle ausgewählt, die die Kontrollverbindung betreffen. Kommandos hingegen, die Dateien herunterladen, hochladen oder Verzeichnisse erstellen, wurden nicht berücksichtigt, da das Antwort-Verhalten von den Dateistrukturen des Zielsystems abhinge und sogar durch den Fingerprinting-Vorgang beeinflusst werden könnte.

Für jeden der ausgesuchten FTP-Befehle und jede mögliche Länge bis zu einer vom Anwender definierten oberen Schranke wurden Argumente erzeugt. So wird nicht nur ein Test auf unterschiedliche Eingabemengen, sondern auch Tests hinsichtlich des Verhaltens bei unterschiedlich langen Argumenten durchgeführt. Von jeder erzeugten Anfrage-Nachricht für einen gegebenen FTP-Befehl und eine gegebene Länge werden n Instanzen erzeugt, die jeweils m-mal mutieren. Beide Parameter können durch den Anwender der Software bestimmt werden. Durch die mehrfache Mutation wird die konkrete Eingabemenge des Fingerprintings vergrößert und damit die Genauigkeit des Fingerprintings potentiell verbessert.

Zur Mutation wird ein Algorithmus verwendet, der die übergebene Nachricht bei jedem Aufruf entweder um ein zufällig gewähltes Zeichen verlängert, zufällig ein Zeichen ändert oder zufällig ein Zeichen aus der Nachricht entfernt. Die Anfrage-Nachrichten werden grundsätzlich nur einmal für jede Versuchsreihe erzeugt. So kann jedes Zielsystem mit der gleichen Eingabemenge getestet werden, um die Vergleichbarkeit der Antwort-Nachrichten zu gewährleisten.

## 4.1 Komponenten der Implementierung

Die Implementierung erfolgte in der Programmiersprache Python auf Basis der Fuzzing-Bibliothek Antiparser [antiparser], als Entwicklungsplattform diente Linux. Grundsätzlich ist die entwickelte Software jedoch auch auf andere Betriebssysteme wie Windows und MacOS portierbar. Die Software besteht i.W. aus 3 Komponenten, dem *Fuzzer*, dem *Sender* und dem *Matcher*.

Der *Fuzzer* erzeugt die Anfragenachrichten und speichert diese in einer Fuzz-Collection. Um die Fingerprints verschiedener Zielsysteme vergleichbar zu halten, werden jeweils die gleichen Anfragenachrichten für das Fingerprinting genutzt. Die Fuzz-Collection ist hierbei eine Textdatei, die alle erzeugten Anfragenachrichten enthält. Dabei wird in jeder Zeile der Datei genau eine Anfragenachricht gespeichert. Der Fuzzer setzt sich aus zwei Modulen zusammen. Zum einen erzeugt die Antiparser-Bibliothek FTP-Befehle mit Argumenten unterschiedlicher Länge, zum anderen erzeugt die Funktion *mutate* verschiedene Instanzen der erzeugten Befehle einer bestimmten Länge. Diese Befehle werden in der Fuzz-Collection gespeichert und beim Fingerprinting eines Hosts wieder abgerufen.

Der *Sender* verschickt die gesammelten Anfragenachrichten an das Zielsystem und speichert die Antwortnachrichten eines Zielsystems in einer Fingerprint-Collection. Der Sender baut hierzu eine TCP-Verbindung zum Zielsystem auf und loggt sich anonym auf dem FTP-Server ein. Sofern für die Tests Zugangsdaten vorliegen und kein anonymer Zugriff auf dem FTP-Zielsystem erlaubt ist, können die Zugangsdaten auch als Parameter übergeben werden. Anschließend werden in einer Schleife abwechselnd Anfragenachrichten aus der Fuzz-Collection versandt, die zugehörige Antwortnachricht empfangen und in der Fingerprint-Collection gespeichert. Zum gegenwärtigen Zeitpunkt speichert die implementierte Service-Fingerprinting-Software lediglich den FTP-Statuscode einer Antwortnachricht in Textdateien.

Der *Matcher* ist die Komponente zum Vergleich eines Fingerprints mit den übrigen Fingerprints der Fingerprint-Collection. Anhand der prozentualen Übereinstimmung zwischen jeweils 2 Fingerprints lassen sich Rückschlüsse über den verwendeten FTP-Server des Zielsystems herstellen. Die Ergebnisse werden also nicht von der Fingerprinting-Software selbst interpretiert. Im Ergebnis wird eine Liste der besten Matches ausgegeben, deren Länge vom Benutzer konfigurierbar ist. Der Matcher iteriert über die Zeilen zweier Fingerprint-Dateien und ermittelt die Übereinstimmung der empfangenen Statuscodes. Dem Anwender werden im Ergebnis die fünf Fingerprints mit dem höchsten Übereinstimmungsgrad gezeigt.

Die wesentlichen Komponenten und deren Interaktion sind in Abbildung 2 dargestellt.

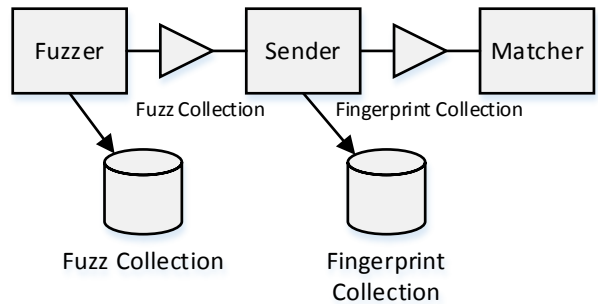

**Abb. 2:** Basiskomponenten des Fingerprinting-Tools

Die entwickelte Fingerprinting-Software ist als Kommandozeilenprogramm realisiert. Um eine erhöhte Benutzerfreundlichkeit zu erzielen, wurde zusätzlich eine grafische Benutzeroberfläche implementiert.

## 5 Ergebnisse

Um die entwickelte Fingerprinting-Software hinsichtlich ihrer Erkennungsgenauigkeit zu testen, wurden 6 verschiedene FTP-Serverprodukte auf einem Testsystem installiert. Die Ergebnisse des Fuzzing-basierten Fingerprintings zeigen deutliche Unterschiede zwischen den verschiedenen FTP-Servern. Die höchste Übereinstimmung zwischen zwei verschiedenen FTP-Serverprodukten lag bei etwa 81% (vgl. Tabelle 1), während unterschiedliche Versionen des gleichen FTP-Serverprodukts erwartungsgemäß eine sehr viel größere Übereinstimmung zeigten, z.T. aber auch nicht unterscheidbar waren.

**Tab. 1:** Übereinstimmungs-Matrix für verschiedene FTP-Serverprodukte.

| FTP-Server | GLFTPD | Net-FTPServer | Netkit-FTPD | PROFTPD | Pure-FTPD | VSFTPD |
|---|---|---|---|---|---|---|
| GLFTPD | 100,00% | 37,82% | 81,41% | 51,54% | 37,92% | 46,15% |
| Net-FTPServer | 37,82% | 100,00% | 38,46% | 56,41% | 74,36% | 58,97% |
| Netkit-FTPD | 81,41% | 38,46% | 100,00% | 51,92% | 43,59% | 39,10% |
| PROFTPD | 51,54% | 56,41% | 51,92% | 100,00% | 51,28% | 51,32% |
| Pure-FTPD | 37,92% | 74,36% | 43,59% | 51,28% | 100,00% | 66,67% |
| VSFTPD | 46,15% | 58,97% | 39,10% | 51,32% | 66,67% | 100,00% |

Auch nach mehreren Durchläufen bleiben diese Ergebnisse konstant. Diese Zahlen erlauben die folgenden Rückschlüsse:

- Das Antwort-Verhalten der verschiedenen FTP-Server ist signifikant unterschiedlich und erlaubt tatsächlich ein Fingerprinting mit den durch Fuzzing erzeugten Daten.
- Stimmen die Fingerprints zweier Server zu 100% überein, so kann davon ausgegangen werden, dass es sich um das gleiche Produkt handelt.

Anzumerken ist, dass die Unterscheidung von FTP-Servern nur dann möglich ist, wenn entweder ein anonymer FTP-Zugang vorhanden ist oder gültige Zugangsdaten verwendet werden. Ist dies nicht der Fall, reagieren alle getesteten FTP-Server mit demselben Fehlercode und können daher nicht differenziert werden. Dieser Einschränkung unterliegen allerdings auch andere Fingerprinting-Tools.

Für den Fall, dass sich die Fingerprints zweier Zielsysteme unterscheiden, sind folgende Rückschlüsse möglich:

- Die Zielsysteme verwenden unterschiedliche FTP-Server. Die prozentuale Übereinstimmung der Fingerprints ist ein guter Indikator dafür, ob es sich um zwei völlig verschiedene Produkte handelt oder nicht. Fraglich ist jedoch, wo die Grenze gezogen werden soll. Wie aus Tabelle 1 ersichtlich, beträgt der höchste Übereinstimmungsgrad zwischen zwei verschiedenen Servern 81,41%. Auf Grundlage oben genannter Daten könnte etwa eine Grenze bei 90% gezogen werden. Um zu einem hinreichend genauen Ergebnis zu kommen, müssten jedoch wesentlich mehr FTP-Server getestet werden.
- Die Zielsysteme nutzen das gleiche FTP-Serverprodukt, verwenden jedoch unterschiedliche Betriebssysteme. Dies kann einen z.T. erheblichen Einfluss auf das Ergebnis haben. In einem Vergleich des Pure-FTPD 1.0.21 unter Linux und Windows (innerhalb einer Cygwin-Umgebung) ergab sich lediglich eine Übereinstimmung von 72,91% beider Fingerprints. Somit erscheint es als sinnvoll, Fingerprints für beide Betriebssysteme vorzuhalten.
- Die Zielsysteme verwenden unterschiedliche Versionen der gleichen Software. So ergibt sich zwischen den Versionen 2.0.5 und 2.0.7 des VSFTPD eine reproduzierbare Verhaltensübereinstimmung von 97,80%. Der Unterschied von 2,2% ist demnach ein hinreichend sicheres Unterscheidungsmerkmal. Die Möglichkeit, einzelne Versionen voneinander zu unterscheiden, ist dennoch begrenzt, da manche Versionssprünge schlicht keine Änderungen im Antwortverhalten bewirken. So konnte etwa mit der konzipierten Lösung kein Verhaltens-Unterschied zwischen den Versionen 1.0.21 und 1.0.11 des Pure-FTPD ausgemacht werden.

## 5.1 Vergleich mit bestehender Fingerprinting-Software

Um die Effektivität des gewählten Fuzzing-basierten Ansatzes mit bestehenden Ansätzen zu vergleichen, wurden Tests mit etablierten FTP-Fingerprinting-Tools vorgenommen.

Vor dem Testen wurde die Banner-Ausgabe jedes FTP-Servers deaktiviert, um die Vorteile eines Fingerprintings gegenüber der Anwendung von Banner-Grabbing zu demonstrieren. Über Banner-Grabbing lassen zwar sich schnell präzise Ergebnisse erreichen, jedoch ist die Deaktivierung oder Änderung des Banners gängige Praxis und fast jeder erhältliche FTP-Server bietet eine entsprechende Einstellung an. Die einzige Ausnahme in diesem Test war der Netkit-FTPD. Um die Ausgabe des Banner-Strings zu deaktivieren, musste daher der Quellcode angepasst und die Server-Software neu kompiliert werden.

Die Fingerprinting-Software THC amap 5.2 [amap] ist ohne Ausgabe von Banner-Strings nutzlos. Trotz der Verwendung von im Vorhinein gespeicherten Erkennungsmerkmalen konnte nach der Deaktivierung der Banner-Strings keiner der getesteten FTP-Server von amap identifiziert werden. Als einzige Information kann die Verwendung des File Transfer Protocol der Ausgabe entnommen werden. Dagegen konnte mit der selbst entwickelten Fingerprinting-Software jede der getesteten Server-Varianten reproduzierbar identifiziert und durch eine prozentuale Übereinstimmung klar von anderen FTP-Servern abgegrenzt werden. Es ist ersichtlich, dass der gewählte Ansatz dem Banner-Grabbing deutlich überlegen ist.

Das Programm FTPmap 0.4 [FTPmap] ist eine dedizierte FTP-Fingerprinting-Software. Es werden Standard-Testfälle aus einer Konfigurationsdatei gelesen, an den entfernten FTP-Server versandt und die Antworten ausgewertet. Nachdem Fingerprints für alle eingesetzten FTP-Server erstellt wurden, konnte FTPmap jeden Dienst bei späteren Tests wiedererkennen. Um eine spezifische Server-Software zu erkennen, ist dieses Vorgehen somit völlig ausreichend.

In einem weiteren Schritt wurde untersucht, inwiefern sich unterschiedliche Versionen eines FTP-Serverprodukts mit FTPmap unterscheiden lassen. Die Vergleichstests umfassten die Installation verschiedener FTP-Serverversionen in der gleichen Systemumgebung und das Fingerprinting dieser Versionen mit FTPmap und der hier entwickelten Fingerprinting-Lösung.

Dabei zeigte sich, dass die selbst entwickelte Fingerprinting-Software im Unterschied zu FTPmap in der Lage ist, gewisse Versionen eines Produkts zu unterscheiden. Im vorliegenden Fall konnte eine Unterscheidung der Versionen 2.0.5 und 2.0.7 von VSFTPD unter Linux 2.6.15 (Gentoo) erreicht werden. Der Verhaltensunterschied betrug dabei reproduzierbar 2,2%. Diese Unterscheidung ist mit FTPmap nicht möglich.

Die weitere Analyse ergab, dass zu FTPmap eine Datei mit Anfragenachrichten gehört, die an das Zielsystem versandt werden. Die Schwäche des Programms liegt letztendlich in diesen manuell gewählten Anfragenachrichten. Somit ist es auch möglich, FTPmap durch Einsatz der mittels Fuzzing generierten Anfragen zu verbessern und den Vorsprung der hier entwickelten Fingerprinting-Software wieder aufzuholen. Dafür extrahiert die von den Verfassern geschaffene Fingerprinting-Lösung genau jene Testfälle, deren Antwortnachrichten sich in verschiedenen Servern und Serverversionen unterscheiden und somit ein Identifizierungsmerkmal bieten.

Dasselbe Vorgehen ist grundsätzlich auch beim HTTP-Fingerprinting und anderen Protokollen möglich. Ein Beispiel für eine auf HTTP-Fingerprinting-Software spezialisierte Software ist hmap [hmap]. Genau wie bei FTPmap werden manuell erzeugte Testfälle verwendet, was den Rückschluss zulässt, dass es ebenfalls Verbesserungspotentiale in der Erkennung von Webservern gibt. Um die von den Verfassern entwickelte Fingerprinting-Software für den Einsatz in HTTP zu benutzen, müssten die FTP-Kommandos durch ihre HTTP-Äquivalente ersetzt werden. Die Reintegration der relevanten Testfälle in hmap kann anschließend auf die gleiche Weise wie bei FTPmap erfolgen. Eine vollständige praktische Erprobung des HTTP-Fingerprinting mittels Fuzzing wurde bisher jedoch nicht im Rahmen der Tests vorgenommen.

# 6 Fazit und Ausblick

Der hier vorgestellte Fuzzing-basierte Ansatz ist in der Lage, zuverlässige Unterscheidungen von Diensten und teilweise sogar deren Versionsständen zu treffen. Dies erfolgt mit der Genauigkeit von State-Of-The-Art Tools, in Teilen sogar mit einer höheren Genauigkeit. Die Referenzimplementierung zeigt, dass sich FTP-Serverprodukte leicht unterscheiden und identifizieren lassen. Auch die Zuordnung zu einer bestimmten Version des jeweiligen FTP-Servers ist teilweise möglich.

Der Aufwand bei der Suche nach neuen Unterscheidungsmerkmalen wird mit dem vorgestellten Verfahren erheblich reduziert und damit die Entwicklung neuer Fingerprinting-Software erleichtert. Für die Anwendung in Penetrationstests können genau jene Anfragenachrichten extrahiert werden, die eine Unterscheidung zwischen zwei Zielsystemen ermöglichen. Dies erhöht die Performance von Penetrationstests.

Der nächste logische Entwicklungsschritt des vorgestellten Konzepts ist die Erweiterung des Fingerprintings um weitere Anwendungsprotokolle wie beispielsweise HTTP und SMTP. Ein weiterer Ansatzpunkt ist die Einbindung weiterer Fuzzing-Methoden, insbesondere des Evolutionary Fuzzings, um die Suche nach den geeigneten Anfragenachrichten zu beschleunigen.

## Literatur

[Lippmann et al.] Lippmann, R.; Fried, D.; Piwowarski, K.; Streilein, W., „Passive operating system identification from TCP/IP packet headers", in Proceedings of the ICDM Workshop on Data Mining for Computer Security, 2003.

[Sarraute und Burroni] Sarraute, Carlos; Burroni, Javier, „Using Neural Networks to improve classical Operating System Fingerprinting techniques", Electronic Journal of SADIO, Vol. 8, No. 1, 2008.

[Gagnon et al.] Gagnon, Francois; Esfandiari, Babak; Bertossi, L., „A Hybrid Approach to Operating System Discovery using Answer Set Programming", Integrated Network Management, IEEE, 2007.

[Caballero et al.] Caballero, J.; Venkataraman, S.; Poosankam, P.; Kang, M. G.; Song, D.; Blum, A., „Automatic fingerprint generation", Network and Distributed System Security Symposium, 2007.

[Greenwald und Thomas] Greenwald, Lloyd G.; Thomas, Tavaris J, „Toward undetected operating system fingerprinting", in Proceedings of the first USENIX workshop on Offensive Technologies, USENIX, 2007.

[amap] amap, http://freeworld.thc.org/thc-amap/

[hmap] hmap, http://ujeni.murkyroc.com/hmap/

[FTPMap] FTPMap, http://linux.softpedia.com/get/Internet/FTP/Ftpmap-26663.shtml

[antiparser] Antiparser-Bibliothek, http://antiparser.sourceforge.net/